\documentclass[eqsecnum,twocolumn,aps,epsf]{revtex4}
\usepackage{graphicx}
\usepackage{color}
\begin{document}
\draft

\title{Muon-spin-relaxation and magnetic-susceptibility studies of the effects of nonmagnetic impurities on the Cu-spin dynamics and superconductivity in La$_{2-x}$Sr$_x$Cu$_{1-y}$Zn$_y$O$_4$ around $x=0.115$}

\author{T. Adachi, S. Yairi, K. Takahashi, Y. Koike}

\address{Department of Applied Physics, Graduate School of Engineering, Tohoku University,\\Aoba-yama 05, Aoba-ku, Sendai 980-8579, Japan}

\author{I. Watanabe}

\address{Muon Science Laboratory, RIKEN (The Institute of Physical and Chemical Research), 2-1 Hirosawa, Wako 351-0198, Japan}

\author{K. Nagamine}

\address{Meson Science Laboratory, Institute of Materials Structure Science, High Energy Accelerator Research Organization, 1-1 Oho, Tsukuba 305-0801, Japan}
\date{\today}

\begin{abstract}

Zero-field muon-spin-relaxation ($\mu$SR) and magnetic-susceptibility measurements have been carried out in La$_{2-x}$Sr$_x$Cu$_{1-y}$Zn$_y$O$_4$ with $x=$ 0.10, 0.115 and 0.13 changing the Zn-concentration $y$ finely up to 0.10, with the aim to clarify effects of the nonmagnetic impurity Zn on the Cu-spin dynamics and superconductivity. 
The $\mu$SR measurements have revealed that, in each $x$, a magnetic order of Cu spins is observed at $y\sim$ 0.0075, while it disappears and Cu spins turn into a fast fluctuating state for $y>$ 0.03. 
From the magnetic-susceptibility measurements, on the other hand, it has been found that the volume fraction of the superconducting state rapidly decreases through the slight doping of Zn and that its $y$ dependence corresponds to the $y$ dependence of the volume fraction of the fast fluctuating region of Cu spins estimated from the $\mu$SR results. 
Both the rapid decrease of the volume fraction of the superconducting state and the formation of the magnetic order through the slight doping of Zn can be interpreted in terms of the development of the so-called "Swiss cheese" model. 
That is, it is concluded that Cu spins in a non-superconducting region around Zn exhibit slowing-down of the fluctuations or form an incoherent or coherent static magnetic order. 
Moreover, the formation of a non-superconducting region around Zn is considered to be due to the pinning of the dynamical spin correlation or dynamical stripe correlations. 

\end{abstract}
\vspace*{2em}
\maketitle
\newpage

\section{Introduction}\label{intro}

There have been a great number of studies of the effects of nonmagnetic impurities in the high-$T_{\rm c}$ superconductors in past years. 
It is well-known that local magnetic moments due to Cu spins are induced by the nonmagnetic impurity Zn in the underdoped regime~\cite{mahajan,xiao} and that the superconductivity is strongly suppressed by Zn.~\cite{maeno,koike} 
Recently, from the scanning-tunneling-microscopy (STM) measurements in the Zn-substituted Bi-2212 system, it has been found that the density of states of quasiparticles increases around each Zn in the superconducting state, suggesting the local destruction of superconductivity.~\cite{pan} 

On the other hand, the effect of nonmagnetic impurities on the dynamical stripe correlations of spins and holes suggested from the neutron scattering measurements in La$_{1.6-x}$Nd$_{0.4}$Sr$_x$CuO$_4$ with $x=0.12$,~\cite{nature,prb} has been another issue with rising interest. 
It has been proposed that the dynamical stripe correlations are pinned and stabilized not only by the characteristic structure (the tetragonal low-temperature structure in the La-214 system) but also by nonmagnetic impurities such as Zn. 
This pinning has been considered to cause the so-called 1/8 anomaly, namely, the conspicuous suppression of superconductivity at $p$ (the hole concentration per Cu) $\sim1/8$. 
In La$_{2-x}$Sr$_x$Cu$_{1-y}$Zn$_y$O$_4$ with $x=0.115$, in fact, the superconductivity is dramatically suppressed through only 1 \% substitution of Zn for Cu.~\cite{koike2} 
Detailed measurements of the thermoelectric power and Hall coefficient in La$_{2-x}$Sr$_x$Cu$_{1-y}$Zn$_y$O$_4$ have suggested that the dynamical stripe correlations tend to be pinned and stabilized by a small amount of Zn, while the static stripe order is destroyed by a large amount of Zn.~\cite{koike3,koike4,adachi} 
Such a stripe-pinning model owing to a small amount of Zn is supported by the theoretical studies based upon the $t$-$J$ model.~\cite{tohyama,smith}

Muon-spin-relaxation ($\mu$SR) measurements have provided a crucial probe to study the magnetism of the 1/8 anomaly in the La-214 system.~\cite{torikai,kumagai,luke,watanabe,watanabe2} 
In La$_{2-x}$Ba$_x$CuO$_4$ with $x=0.125$, for instance, muon-spin precession has been observed at low temperatures below $\sim30$ K, corresponding to the formation of a magnetic order of Cu spins.~\cite{kumagai,luke,watanabe} 
Moreover, the $\mu$SR technique makes it possible to distinguish between magnetic and nonmagnetic regions. 
A $\mu$SR study focusing on the Zn effect has been reported by Nachumi {\it et al.} for $y\le0.01$ in La$_{2-x}$Sr$_x$Cu$_{1-y}$Zn$_y$O$_4$ with $x=0.15$ and 0.20 and for $y\le0.015$ in YBa$_2$(Cu$_{1-y}$Zn$_y$)$_3$O$_{6.63}$, where the so-called "Swiss cheese" model has been proposed.~\cite{nachumi} 
They have observed two kinds of signal due to magnetic and nonmagnetic regions in one sample in the transverse-field (TF) $\mu$SR measurements, so that they have speculated that each Zn introduced in the CuO$_2$ plane produces a non-superconducting region around itself like Swiss cheese. 
However, this model has given no information about the Cu-spin fluctuations. 

In this paper, we investigate the Cu-spin dynamics from the zero-field (ZF) $\mu$SR measurements of the Zn-substituted La$_{2-x}$Sr$_x$Cu$_{1-y}$Zn$_y$O$_4$ with $x=0.10$, 0.115 and 0.13 changing $y$ finely up to 0.10. 
We aim at clarifying from the viewpoint of the spin dynamics whether the pinning of the dynamical stripe correlations by Zn is observed or not. 
We evaluate the Zn-concentration dependence of the volume fractions of different states of Cu spins with very different fluctuation frequencies.~\cite{watanabe3,watanabe4,watanabe5,adachi2} 
We also investigate the Zn effect on the superconductivity from the magnetic-susceptibility measurements. 
We evaluate the Zn-concentration dependence of the volume fraction of the superconducting state. 
Then, in comparison between the volume fractions of different states of Cu spins and the volume fraction of the superconducting state, we aim at elucidating the relation between the Cu-spin dynamics and superconductivity.~\cite{adachi2} 

\section{Experimental details}

Polycrystalline samples of La$_{2-x}$Sr$_x$Cu$_{1-y}$Zn$_y$O$_4$ with $x=0.10$, 0.115, 0.13 and $y=0$, 0.0025, 0.005, 0.0075, 0.01, 0.02, 0.03, 0.05, 0.07, 0.10 were prepared by the ordinary solid-state reaction method. 
Appropriate powders of dried La$_2$O$_3$, SrCO$_3$, CuO and ZnO were mixed and prefired in air at 900$^o$C for 12 h. 
The prefired materials were mixed and pelletized, followed by sintering in air at 1050$^o$C for 24 h. 
Post annealing was performed in flowing oxygen gas at 500$^o$C for 72 h in order to keep oxygen deficiencies to a minimum. 
All of the samples were checked by the powder X-ray diffraction measurements to be single phase. 
The electrical resistivity was also measured to check the quality of the samples. 

The ZF-$\mu$SR measurements were performed at the RIKEN-RAL Muon Facility at the Rutherford-Appleton Laboratory in the UK, using a pulsed positive surface muon beam with an incident muon momentum of 27 MeV/c. 
The asymmetry parameter $A(t)$ at a time $t$ was given by $A(t)=\{F(t)-\alpha B(t)\}/\{F(t)+\alpha B(t)\}$, where $F(t)$ and $B(t)$ were total muon events of the forward and backward counters, which were aligned in the beam line, respectively. 
The $\alpha$ is the calibration factor reflecting the relative counting efficiencies between the forward and backward counters. 
The $\mu$SR time spectrum, namely, the time evolution of $A(t)$ was measured down to 2 K to detect the appearance of a magnetic order. 

Magnetic-susceptibility measurements were carried out down to 2 K using a standard SQUID magnetometer (Quantum Design, Model MPMS-XL5) in a magnetic field of 10 Oe on field cooling, in order to evaluate the volume fraction of the superconducting state. 
In the case that a non-superconducting region in a sample is surrounded by a superconducting region, in fact, the superconducting volume fraction should be estimated from the value of the magnetic susceptibility on field cooling rather than on zero-field cooling. 
Even on field cooling, the vortex pinning effect in the superconducting region must be taken into account to evaluate the volume fraction precisely, but the effect is negligible to evaluate the volume fraction roughly in La$_{2-x}$Sr$_x$CuO$_4$.~\cite{yairi,nagano}

\section{Results}
\subsection{ZF-$\mu$SR}\label{sec:A}
\begin{figure*}[tbp]
\begin{center}
\includegraphics[width=0.51\linewidth]{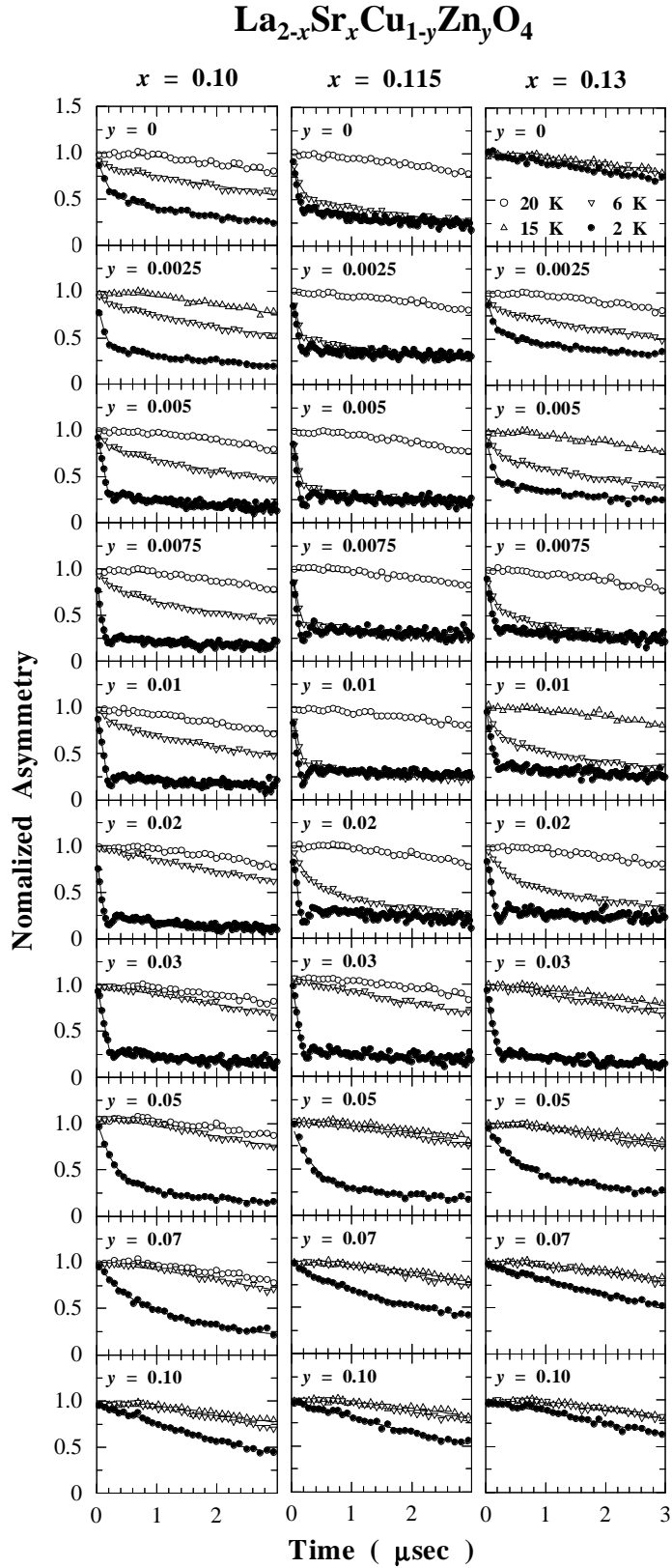}
\end{center}
\caption{ZF-$\mu$SR time spectra of La$_{2-x}$Sr$_x$Cu$_{1-y}$Zn$_y$O$_4$ with $x=0.10$, 0.115, 0.13 and $y=0 - 0.10$ at various temperatures down to 2 K. Time spectra in the early time region from 0 to 3 $\mu$sec are displayed. Solid lines indicate the best-fit results using $A(t) = A_0 e^{-\lambda_0t}G_Z(\Delta,t) + A_1 e^{-\lambda_1t} + A_2 e^{-\lambda_2t}{\rm cos}(\omega t + \phi)$.}  
\label{fig:spec} 
\end{figure*}

\begin{figure}[tbp]
\begin{center}
\includegraphics[width=0.8\linewidth]{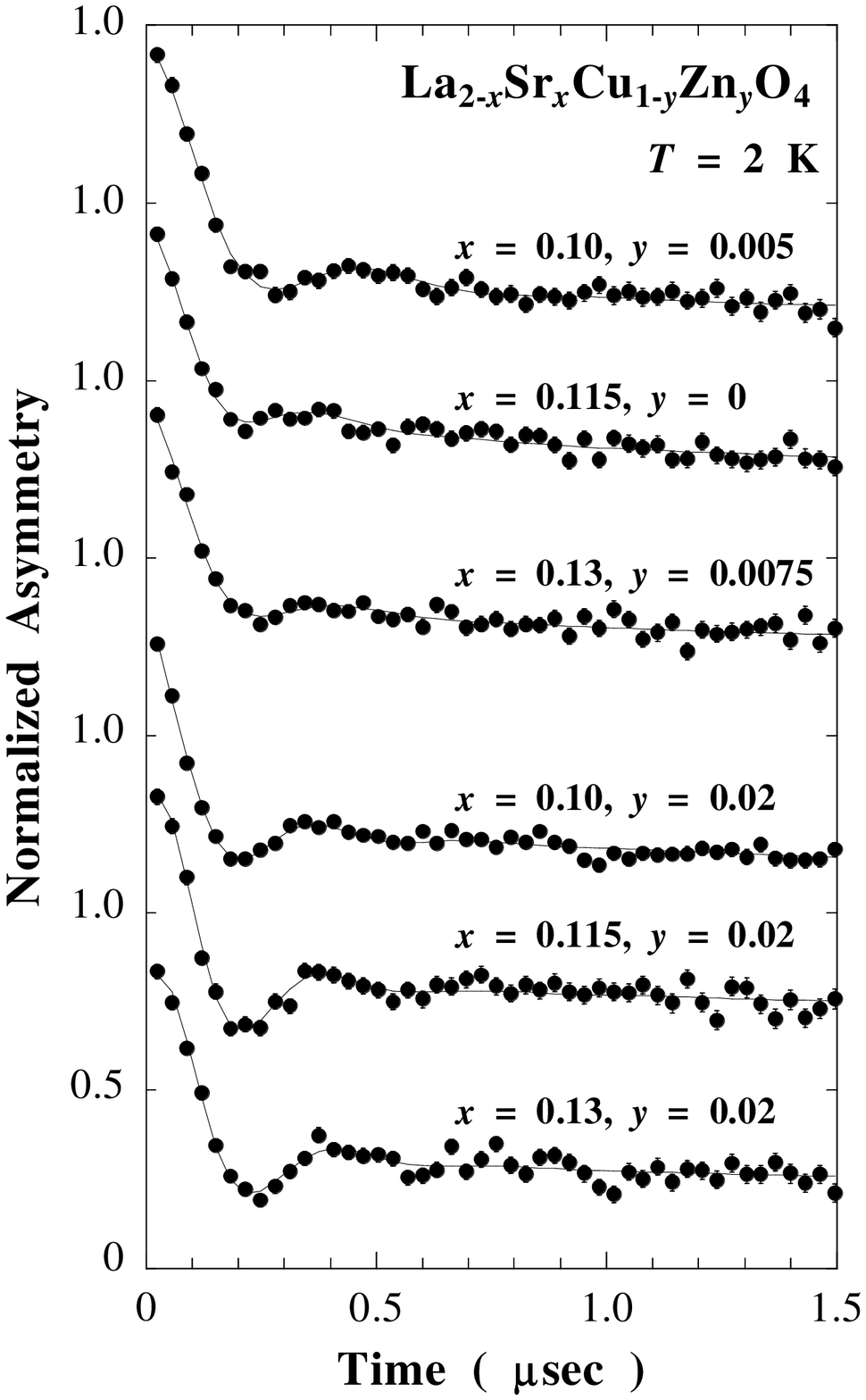}
\end{center}
\caption{ZF-$\mu$SR time spectra at 2 K in the very early time region from 0 to 1.5 $\mu$sec for La$_{2-x}$Sr$_x$Cu$_{1-y}$Zn$_y$O$_4$ with $x=0.10$, 0.115, and 0.13. The $y$ values shown in the upper three spectra correspond to the values where the muon-spin precession starts to be observed with increasing $y$, while those shown in the lower three spectra correspond to the values where the muon-spin precession is most clearly observed among all $y$ in each $x$. Solid lines indicate the best-fit results using $A(t) = A_0 e^{-\lambda_0t}G_Z(\Delta,t) + A_1 e^{-\lambda_1t} + A_2 e^{-\lambda_2t}{\rm cos}(\omega t + \phi)$.} 
\label{fig:spec2} 
\end{figure}

Figure \ref{fig:spec} displays the ZF-$\mu$SR time spectra of La$_{2-x}$Sr$_x$Cu$_{1-y}$Zn$_y$O$_4$ with $x=0.10$, 0.115, 0.13 and $y=0 - 0.10$. 
At a high temperature of 15 K or 20 K, all the spectra show Gaussian-like depolarization due to the randomly oriented nuclear spins. 
Focusing on the spectra at 2 K, muon-spin precession is observed in the Zn-free sample with $x=0.115$. 
This is clearly seen in Fig. \ref{fig:spec2}, indicating the formation of a long-range magnetic order of Cu spins at low temperatures. 
As seen in Fig. \ref{fig:spec}, the deviation of $x$ from 0.115 in the Zn-free samples leads to the destabilization of the magnetic order. 
That is, no precession of muon spins is observed for $x=0.10$. 
Even fast depolarization of muon spins is not observed for $x=0.13$. 
These indicate that the magnetic order is most developed at $x=0.115$ where the superconductivity is most suppressed.~\cite{torikai,watanabe2} 

With increasing $y$, muon-spin precession is clearly observed at $y\sim0.0075$ in each $x$. 
Since the precession pattern is almost identical with one another in each $x$ for $y=0.02$ where the most coherent precession is observed among all $y$, as seen in Fig. \ref{fig:spec2}, it appears that nearly the same magnetic order is formed at $y\sim0.02$ in each $x$. 
For $y>0.03$, on the other hand, the muon-spin precession disappears and almost no fast depolarization of muon spins is observed at $y=0.10$. 
These mean that the static magnetic order is destroyed and that Cu spins are fluctuating with shorter periods than the $\mu$SR time window (10$^{-6} - 10^{-11}$ sec) at $y=0.10$. 

In order to derive the detailed information about the Cu-spin dynamics from the $\mu$SR results, we attempt to analyze the time spectra with the following three-component function: 

\begin{equation}
A(t) = A_0 e^{-\lambda_0t}G_Z(\Delta,t) + A_1 e^{-\lambda_1t} + A_2 e^{-\lambda_2t}{\rm cos}(\omega t + \phi).
\label{eq1}
\end{equation}
The first term represents the slowly depolarizing component in a region where the Cu spins are fluctuating fast beyond the $\mu$SR time window. 
The $A_0$ and $\lambda_0$ are the initial asymmetry and depolarization rate of the slowly depolarizing component, respectively. 
The $G_Z(\Delta,t)$ is the static Kubo-Toyabe function with a half-width of $\Delta$ describing the distribution of the nuclear-dipole field at the muon site.~\cite{uemura} 
The second term represents the fast depolarizing component in a region where the Cu-spin fluctuations slow down and/or an incoherent magnetic order is formed. 
The $A_1$ and $\lambda_1$ are the initial asymmetry and depolarization rate of the fast depolarizing component, respectively.
The third term represents the muon-spin precession in a region where a coherent static magnetic order of Cu spins is formed. 
The $A_2$ is the initial asymmetry. 
The $\lambda_2$, $\omega$ and $\phi$ are the damping rate, frequency and phase of the muon-spin precession, respectively. 
The time spectra are well fitted with Eq. (\ref{eq1}), as shown by solid lines in Fig. \ref{fig:spec} and Fig. \ref{fig:spec2}. 

\begin{figure}[tbp]
\begin{center}
\includegraphics[width=0.8\linewidth]{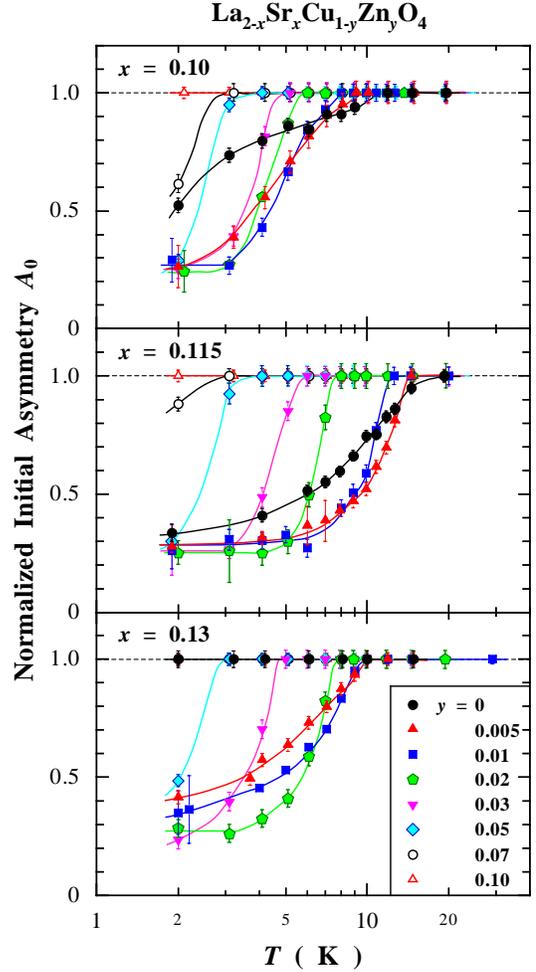}
\end{center}
\caption{(color) Temperature dependence of the initial asymmetry of the slowly depolarizing component $A_0$, normalized by its value at a high temperature of 15 K or 20 K, for the typical $y$ values in La$_{2-x}$Sr$_x$Cu$_{1-y}$Zn$_y$O$_4$ with $x=0.10$, 0.115 and 0.13. Solid lines are to guide the reader's eye.}  
\label{fig:a0} 
\end{figure}

Figure \ref{fig:a0} shows the temperature dependence of $A_0$ obtained from the best-fit using Eq. (\ref{eq1}) for typical $y$ values in La$_{2-x}$Sr$_x$Cu$_{1-y}$Zn$_y$O$_4$ with $x=0.10$, 0.115 and 0.13. 
The value of $A_0$ is normalized by its value at a high temperature of 15 K or 20 K. 
The temperature dependence of $A_0$ is often used as a probe of the magnetic transition, because it reflects the volume fraction of the nonmagnetic region.~\cite{torikai,watanabe,watanabe2}
In fact, $A_0 = 1$ means that the spectrum is represented only with the first term of Eq. (\ref{eq1}), indicating that all the Cu spins are fluctuating fast beyond the $\mu$SR time window. 
On the contrary, $A_0=1/3$ indicates that all the Cu spins are nearly or completely ordered statically. 
As shown in Fig. \ref{fig:a0}, $A_0$ becomes almost 1/3 at 2 K for $y=0.005-0.05$ in $x=0.10$, $y=0-0.05$ in $x=0.115$ and $y=0.01-0.03$ in $x=0.13$, indicating the formation of the magnetic order. 
For the Zn-free samples with $x=0.10$ and 0.115 and the 0.5 \% Zn-substituted sample with $x=0.13$, the transition to the magnetically ordered state is rather broad, though the onset temperature, $T_{\rm N}^{\rm onset}$, defined as the temperature where $A_0$ starts to deviate from 1 with decreasing temperature, is higher than $T_{\rm N}^{\rm onset}$ for the Zn-substituted samples with $x=0.10$ and 0.115 and for the Zn-substituted samples with $x=0.13$ and $y\ge0.02$, respectively. 
With increasing $y$, the magnetic transition tends to be sharp. 
At $y=0.10$, no decrease in $A_0$ from 1 is observed.  
The details of these behaviors depending on $y$ are discussed in Sec. \ref{discussion}.

\begin{figure}[tbp]
\begin{center}
\includegraphics[width=1.0\linewidth]{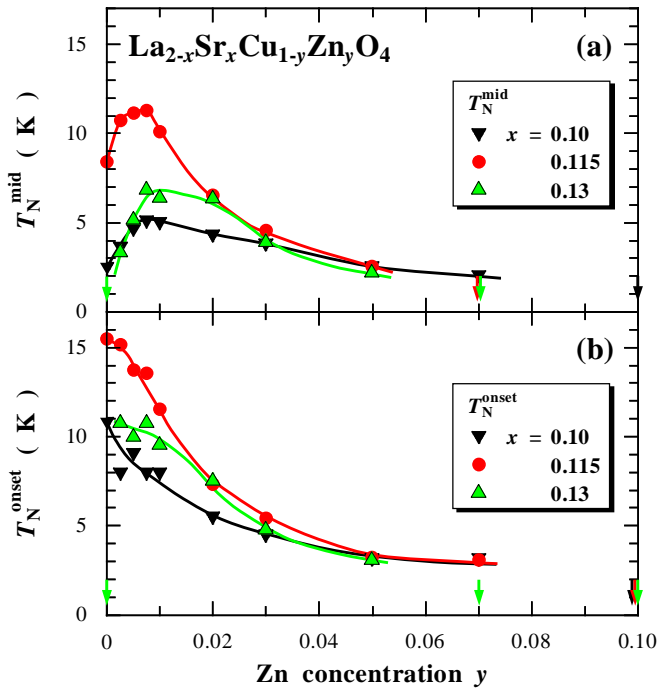}
\end{center}
\caption{(color) Zn-concentration dependence of (a) $T_{\rm N}^{\rm mid}$ defined as the midpoint of the change of the $A_0$ value from 1 to $A_{0_{{\rm min}}}$ and (b) the onset temperature of the magnetic transition, $T_{\rm N}^{\rm onset}$, for La$_{2-x}$Sr$_x$Cu$_{1-y}$Zn$_y$O$_4$ with $x=0.10$, 0.115 and 0.13. Solid lines are to guide the reader's eye.}
\label{fig:tn}
\end{figure}
We estimate the magnetic transition temperature, $T_{\rm N}^{\rm mid}$, defined as the midpoint of the change of the $A_0$ value from 1 to the minimum value $A_{0_{{\rm min}}}$ in each $x$.~\cite{tn} 
Figure \ref{fig:tn}(a) shows the Zn-concentration dependence of $T_{\rm N}^{\rm mid}$ for La$_{2-x}$Sr$_x$Cu$_{1-y}$Zn$_y$O$_4$ with $x=0.10$, 0.115 and 0.13. 
For $x=0.10$ and 0.115, $T_{\rm N}^{\rm mid}$ increases with increasing $y$ and shows the maximum at $y\sim0.0075$. 
In the case of $x=0.13$, $T_{\rm N}^{\rm mid}$ appears at $y=0.0025$ and shows the maximum at $y\sim0.01$. 
Above $y\sim0.01$, on the other hand, $T_{\rm N}^{\rm mid}$ decreases gradually with increasing $y$ and finally disappears above $y\sim0.05$ or 0.07 in each $x$. 
Figure \ref{fig:tn}(b) shows the Zn-concentration dependence of $T_{\rm N}^{\rm onset}$. 
The $T_{\rm N}^{\rm onset}$ decreases almost monotonically with increasing $y$. 
These results indicate that the transition toward the formation of the magnetic order is rather broad at $y\sim0$, while it becomes sharp through the introduction of Zn, as mentioned in the preceding paragraph. 

Next, we estimate the volume fractions of the three different states of Cu spins in one sample related to the three terms in Eq. (\ref{eq1}) respectively, using the best-fit values of $A_0$, $A_1$ and $A_2$. 
As $A_0$ is not zero but $A_{0_{{\rm min}}}$ even in a completely magnetically-ordered state, the volume fraction $V_{\rm A_0}$ of the region where the Cu spins are fluctuating fast beyond the $\mu$SR time window (We call this region the $A_0$ region) is given by

\begin{equation}
V_{\rm A_0} = \frac{A_0-A_{0_{{\rm min}}}}{(A_0 + A_1 + A_2)-A_{0_{{\rm min}}}}.
\label{eq2}
\end{equation}
The volume fraction $V_{\rm A_1}$ of the region where the Cu-spin fluctuations slow down and/or an incoherent magnetic order is formed (the $A_1$ region) is given by

\begin{equation}
V_{\rm A_1} = \frac{A_1}{(A_0 + A_1 + A_2)-A_{0_{{\rm min}}}}.
\label{eq3}
\end{equation}
~\cite{a1} The volume fraction $V_{\rm A_2}$ of the region where a coherent static magnetic order is formed (the $A_2$ region) is given by

\begin{equation}
V_{\rm A_2} = \frac{A_2}{(A_0 + A_1 + A_2)-A_{0_{{\rm min}}}}.
\label{eq4}
\end{equation}

\begin{figure*}[tbp]
\begin{center}
\includegraphics[width=0.8\linewidth]{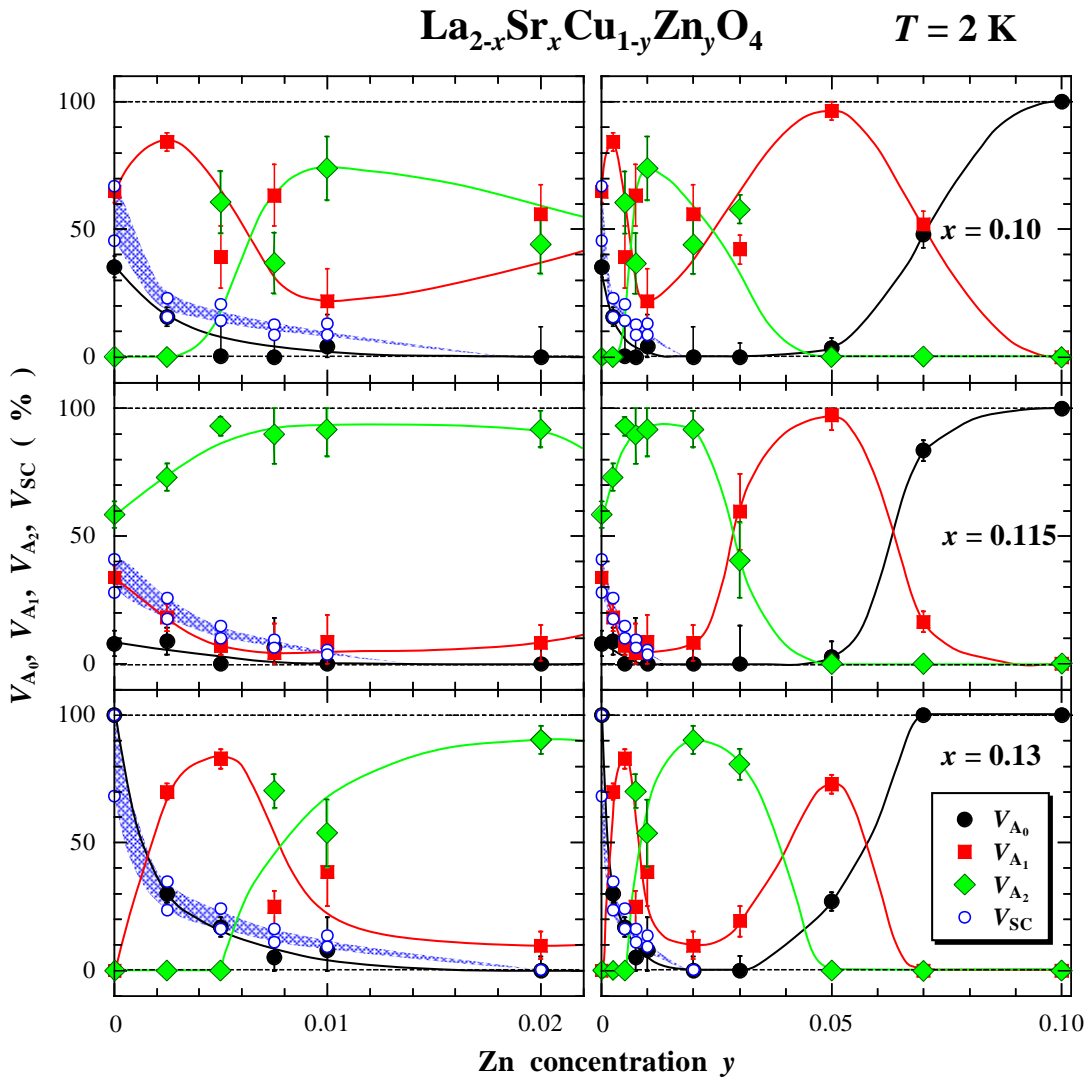}
\end{center}
\caption{(color) Dependence on $y$ of $V_{\rm A_0}$ (black circles), $V_{\rm A_1}$ (red squares) and $V_{\rm A_2}$ (green diamonds) estimated from the $\mu$SR measurements in La$_{2-x}$Sr$_x$Cu$_{1-y}$Zn$_y$O$_4$ with $x=0.10$, 0.115 and 0.13. Dependence on $y$ of $V_{\rm SC}$ (blue shadow area surrounded by blue circles) estimated from the magnetic-susceptibility measurements is also plotted. $V_{\rm A_0}$: Volume fraction of the region where the Cu spins are fluctuating fast beyond the $\mu$SR time window. $V_{\rm A_1}$: Volume fraction of the region where the Cu-spin fluctuation slow down and/or an incoherent magnetic order is formed. $V_{\rm A_2}$: Volume fraction of the region where a coherent static magnetic order is formed. $V_{\rm SC}$: Volume fraction of the superconducting region. Note that left panels are magnification of right panels from $y=0$ to 0.022. Solid lines are to guide the reader's eye.}  
\label{fig:vf} 
\end{figure*}
Figure \ref{fig:vf} displays the Zn-concentration dependence of the volume fraction $V_{\rm A_0}$, $V_{\rm A_1}$ and $V_{\rm A_2}$ at 2 K for La$_{2-x}$Sr$_x$Cu$_{1-y}$Zn$_y$O$_4$ with $x=0.10$, 0.115 and 0.13. 
For $x=0.13$, $V_{\rm A_0}$ is 100 \% at $y=0$, indicating that all the Cu spins are fluctuating at high frequencies. 
With increasing $y$, $V_{\rm A_0}$ rapidly decreases and $V_{\rm A_1}$ increases complementarily, indicating the slowing down of the Cu-spin fluctuations and/or the formation of an incoherent magnetic order. 
In fact, $V_{\rm A_2}$ increases instead of $V_{\rm A_1}$ for $y>0.005$. 
At $y=0.02$, $V_{\rm A_0}$ becomes zero and almost all the Cu spins become static and are magnetically ordered. 
On the other hand, $V_{\rm A_2}$ decreases with increasing $y$ for $y>0.02$, meaning the destabilization of the magnetic order. 
At $y=0.10$, $V_{\rm A_0}$ again becomes 100 \%, indicating that all the Cu spins are fluctuating at high frequencies again. 

As for $x=0.10$ and 0.115, almost similar Zn-concentration dependence of the volume fractions of the respective Cu-spin states are obtained, but the important difference among $x=0.10$, 0.115 and 0.13 is in the Cu-spin state of the Zn-free sample. 
For $x=0.115$, $V_{\rm A_0}$ is less than 10 \%, while $V_{\rm A_2}$ is more than 50 \%. 
For $x=0.10$, on the other hand, $V_{\rm A_1}$ is more than 60 \%, while $V_{\rm A_2}$ is zero. 
These indicate that the magnetic order is most developed at $x=0.115$ in the Zn-free samples and is suddenly destabilized with the deviation of $x$ from 0.115.~\cite{torikai,watanabe2} 
These results also give two important features; (i) the fast fluctuating region of Cu spins is rapidly diminished by the introduction of a small amount of Zn, and (ii) the slowly fluctuating and/or incoherent magnetically-ordered region of Cu spins are replaced by a coherent magnetically-ordered one for $y>0.005$. 
These behaviors are discussed in detail in Sec. \ref{discussion}. 

\subsection{Magnetic susceptibility}
\begin{figure}[tbp]
\begin{center}
\includegraphics[width=1.0\linewidth]{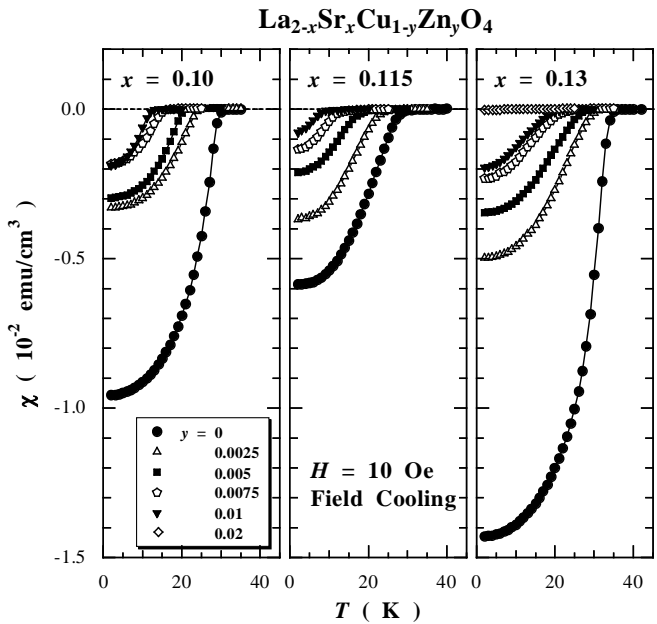}
\end{center}
\caption{Temperature dependence of the magnetic susceptibility $\chi$ of La$_{2-x}$Sr$_x$Cu$_{1-y}$Zn$_y$O$_4$ with $x=0.10$, 0.115 and 0.13 in a magnetic field of 10 Oe on field cooling.}  
\label{fig:chi} 
\end{figure}
Figure \ref{fig:chi} shows the temperature dependence of the magnetic susceptibility $\chi$ for La$_{2-x}$Sr$_x$Cu$_{1-y}$Zn$_y$O$_4$ with $x=0.10$, 0.115, 0.13 and $y\le0.02$. 
In comparison between the absolute values of $\chi$ at 2 K, $|\chi_{\rm 2 K}|$, of the Zn-free samples with $x=0.10$, 0.115 and 0.13, $|\chi_{\rm 2 K}|$ in $x=0.115$ is less than half of that in $x=0.13$. 
This means that the volume fraction of the superconducting state for $x=0.115$ is markedly suppressed even in the Zn-free sample. 
For $x=0.13$, on the other hand, $|\chi_{\rm 2 K}|$ dramatically decreases through only 0.25 \% substitution of Zn. 
With increasing $y$ from $y=0.0025$, $|\chi_{\rm 2 K}|$ decreases gradually and is almost zero at $y=0.02$. 
This peculiar behavior of the decrease of $|\chi_{\rm 2 K}|$ with increasing $y$ suggests that the volume fraction of the superconducting state does not decrease linearly but is strongly suppressed by a very small amount of Zn. 
It is noted that the $y$ dependence of $|\chi_{\rm 2 K}|$ for $y\ge0.0025$ in each $x$ is roughly similar to one another, suggesting that the Zn-substituted states for $y\ge0.0025$ are not so dependent on $x$. 

We estimate the volume fraction of the superconducting state, $V_{\rm SC}$, concretely from the value of $|\chi_{\rm 2 K}|$. 
The value of $|\chi_{\rm 2 K}|$ corresponding to 100 \% of $V_{\rm SC}$ is hard to be determined exactly. 
Thus, we choose two values, namely, the values of $|\chi_{\rm 2 K}|$ of the Zn-free sample with $x=0.13$ and of the optimally doped Zn-free sample with $x=0.15$ as corresponding to 100 \% of $V_{\rm SC}$.~\cite{yairi} 
Then, $V_{\rm SC}$ is estimated as the value of $|\chi_{\rm 2 K}|$ in each $x$ and $y$ divided by that of the Zn-free sample with $x=0.13$ or 0.15, as shown by blue circles in Fig. \ref{fig:chi}. 
The true $y$ dependence of $V_{\rm SC}$ must be located in the shadow area surrounded by the blue circles. 
It is found that the $y$ dependence of $V_{\rm SC}$ is analogous to that of $V_{\rm A_0}$. 
Especially for $x=0.13$, the $y$ dependence of $V_{\rm SC}$ is in good agreement with that of $V_{\rm A_0}$. 
This gives an important information beyond the "Swiss cheese" model. 
That is, the superconductivity is realized in the region where Cu spins fluctuate fast beyond the $\mu$SR time window. 
In other words, Cu spins in a non-superconducting region around Zn exhibit slowing-down of the fluctuations or form an incoherent or coherent static order. 

\section{Discussion}\label{discussion}
\begin{figure}[tbp]
\begin{center}
\includegraphics[width=1.0\linewidth]{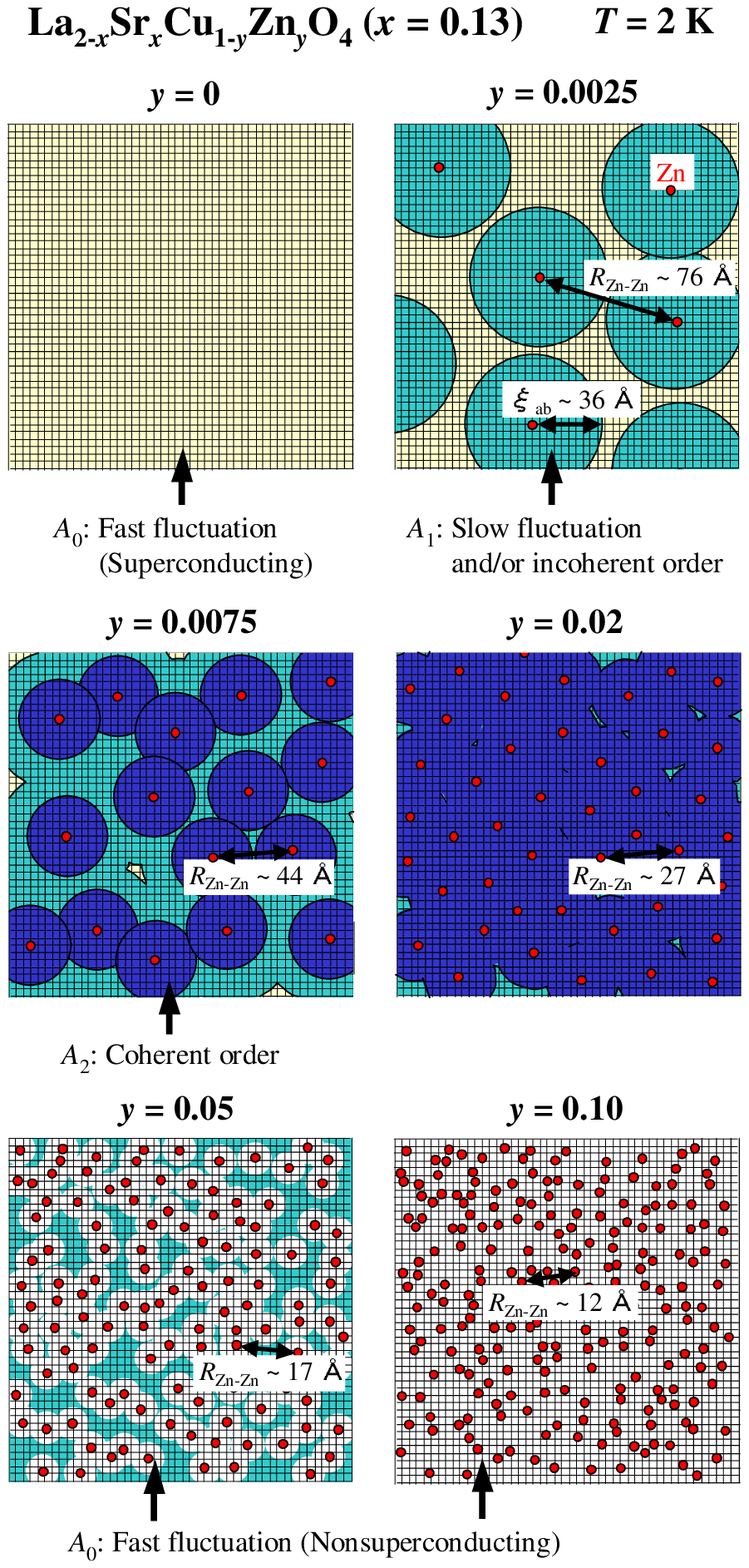}
\end{center}
\caption{(color) Schematic pictures of the spatial distribution of the different Cu-spin states in the CuO$_2$ plane, corresponding to $A_0$, $A_1$ and $A_2$, at $T = 2$ K for typical $y$ values in La$_{2-x}$Sr$_x$Cu$_{1-y}$Zn$_y$O$_4$ with $x=0.13$. Each crossing point of the grid pattern represents the Cu site. Zn atoms are randomly distributed in the CuO$_2$ plane. Sizes of the circles are calculated from the ratio of $V_{\rm A_0}$: $V_{\rm A_1}$: $V_{\rm A_2}$ in each $y$. The $\xi_{\rm ab}$ indicates the radius of the $A_1$ region around each Zn. The $R_{\rm Zn-Zn}$ indicates a mean distance between Zn atoms.}  
\label{fig:model} 
\end{figure}

\begin{figure}[tbp]
\begin{center}
\includegraphics[width=0.9\linewidth]{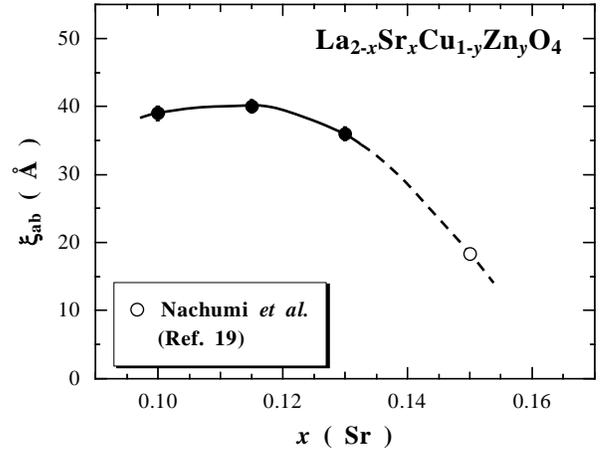}
\end{center}
\caption{Dependence on $x$ of the radius $\xi_{{\rm ab}}$ of the $A_1$ region around each Zn, estimated from the ratio of $V_{\rm A_0}$: $V_{\rm A_1}$: $V_{\rm A_2}$ at $y=0.0025$, for La$_{2-x}$Sr$_x$Cu$_{1-y}$Zn$_y$O$_4$. The data of $x=0.15$ are from Ref. [19]. The solid and dashed line is to guide the reader's eye.}  
\label{fig:xi} 
\end{figure}

First, we discuss the Zn-concentration dependence of $V_{\rm A_0}$, $V_{\rm A_1}$, $V_{\rm A_2}$ and $V_{\rm SC}$. 
Figure \ref{fig:model} displays schematic pictures of the spatial distribution of $A_0$, $A_1$ and $A_2$ regions with different Cu-spin states for typical $y$ values in La$_{2-x}$Sr$_x$Cu$_{1-y}$Zn$_y$O$_4$ with $x=0.13$, on the assumption that the magnetic order develops around each Zn for lightly Zn substituted samples.~\cite{3d}
This assumption has not yet been confirmed directly but is reasonable, referring to the result of the STM measurements in the Zn-substituted Bi-2212 system.~\cite{pan} 
Sizes of the circles are calculated from the ratio of $V_{\rm A_0}$: $V_{\rm A_1}$: $V_{\rm A_2}$ in each $y$. 
Through only 0.25 \% substitution of Zn for Cu, an $A_1$ region, namely, a slowly fluctuating and/or incoherent magnetically-ordered region of Cu spins is formed with $\xi_{\rm ab}$, defined as the radius of $A_1$ region at $T = 2$ K, $\sim36$ ${\rm \AA}$ around each Zn in the $A_0$ region, namely, in the fast fluctuating sea of Cu spins.~\cite{xiab} 
It is found that a large part of the CuO$_2$ plane is covered with $A_1$ regions so that $V_{\rm SC}$ in good correspondence to $V_{\rm A_0}$ is strongly diminished. 
Comparing $\xi_{\rm ab}$ with a mean distance between Zn atoms $R_{\rm Zn-Zn}\sim76$ ${\rm \AA}$ at $y=0.0025$, $A_1$ regions are averagely considered not to overlap one another. 
With increasing $y$, $A_1$ regions overlap one another at $y=0.005$, leading to the development of $A_2$ regions, namely, coherent magnetically ordered regions. 
In fact, $A_0$ regions almost disappear at $y=0.0075$ and alternatively $A_2$ regions are formed around Zn. 
Although plural patterns of the spatial distribution of $A_2$ regions are possible, the pattern where $A_2$ regions develop around Zn is shown in Fig. \ref{fig:model}. 
When $y$ reaches 0.02, $A_2$ regions cover the almost whole CuO$_2$ plane, leading to the formation of a nearly coherent magnetic order. 
With increasing $y$ from $y=0.02$, $A_2$ regions decrease. 
They completely disappear at $y=0.05$ and alternatively $A_0$ regions reappear. 
Here, reappeared $A_0$ regions are non-superconducting and considered to be formed around each Zn as shown in Fig. \ref{fig:model}, because a large amount of nonmagnetic Zn destroys the magnetic order of Cu spins. 
At $y=0.10$, the non-superconducting $A_0$ regions cover the whole CuO$_2$ plane. 
These changes with increasing $y$ suggest that the superconducting region competes with both the slowly fluctuating and magnetically ordered regions of Cu spins. 
As for the 0.25 \% Zn-substituted samples with $x=0.10$, 0.115 and 0.13, $\xi_{\rm ab}$ is not so different from one another but largest at $x=0.115$, as shown in Fig. \ref{fig:xi}. 
For $x=0.115$, moreover, $A_2$ regions are developed even for $y\le0.0025$, as show in Fig. \ref{fig:vf}. 
These suggest that the magnetic order tends to develop especially at $x=0.115$. 

Here, we discuss the Cu-spin state of the non-superconducting region around Zn. 
An analogous model on the Zn effect has been proposed by Nachumi {\it et al.}, which is known as the "Swiss cheese" model.~\cite{nachumi} 
In this model, it has been estimated that Zn destroys the superconductivity around itself with the radius $\xi_{\rm ab}=18.3$ ${\rm \AA}$ for $x=0.15$ in La$_{2-x}$Sr$_x$Cu$_{1-y}$Zn$_y$O$_4$. 
Our estimation of $\xi_{\rm ab}\sim40$ ${\rm \AA}$ for $x=0.115$ is more than twice as large as that for $x=0.15$, as shown in Fig. \ref{fig:xi}. 
One may easily guess that the difference in $\xi_{\rm ab}$ originates from that in the hole concentration. 
That is, more holes in $x=0.15$ than in $x=0.115$ are expected to disturb the formation of the static magnetic order. 
However, this may not be the case, because $\xi_{\rm ab}$ for $x=0.10$ is not larger than that for $x=0.115$. 
We speculate that the difference in $\xi_{\rm ab}$ is related to the spin-correlation length in the region where Zn destroys the superconductivity, because $V_{\rm SC}$ is analogous to $V_{\rm A_0}$. 
Considering that the correlation length of the incommensurate dynamical spin correlation estimated from the inelastic neutron scattering measurements~\cite{yamada} is larger for $x=0.12$ than for $x=0.15$, the large non-superconducting region induced by Zn is understood to be due to the large correlation length of the dynamical spin correlation. 
This means that the dynamical spin correlation or the dynamical stripe correlations tend to be pinned and stabilized by Zn. 
Furthermore, the overlap between the pinned regions leads to the formation of the static spin order. 
Although the spin state of the non-superconducting region around each Zn has not been clear in the "Swiss cheese" model, it is identified from our results that the non-superconducting region around each Zn corresponds to the statically ordered region pinned by Zn. 
This interpretation is regarded as a kind of development of the "Swiss cheese" model. 

Comparison between the results of the recent neutron scattering measurements and the present $\mu$SR results is of much interest. 
In La$_{2-x}$Sr$_x$CuO$_4$ with $x=0.10$, Lake {\it et al.}~\cite{lake} have reported that the intensity of the incommensurate elastic magnetic peaks around ($\pi$, $\pi$) in the reciprocal space observed in zero field~\cite{matsushita} is enhanced by the application of magnetic field. 
In the Zn-substituted La$_{2-x}$Sr$_x$Cu$_{1-y}$Zn$_y$O$_4$ with $x=0.15$ where a gap opens below $\sim3.5$ meV in the spin excitation spectrum,~\cite{yamada2} it has been found by Kimura {\it et al.}~\cite{kimura} that Zn induces the so-called in-gap state in the spin gap, suggesting tendency toward the formation of a static spin order. 
These experimental results have been interpreted as induction of a local antiferromagnetic (AF) order inside the vortex core or around Zn. 
Moreover, the induced AF order turns out to have coherency between them with increasing magnetic field or Zn concentration, leading to the enhancement of the incommensurate magnetic peaks. 
Our interpretation of the present $\mu$SR results is quite similar with respect to the formation of a local magnetic order around Zn. 
That is, it is confirmed in Fig. \ref{fig:model} that the overlap between $A_1$ regions around Zn generates coherency between the Cu spins inside $A_1$ regions, leading to the formation of the coherent magnetic order. 
In the case of $x=0.115$ where the magnetic order is formed even in the Zn-free sample, the coherent magnetic order develops a little with increasing Zn concentration. 
This $\mu$SR result is quite consistent with a little enhancement of the incommensurate magnetic peaks induced by magnetic fields in the neutron scattering measurements in La$_{2-x}$Sr$_x$CuO$_4$ with $x=0.12$ by Katano {\it et al.}~\cite{katano} 
In addition, it is noted that $V_{\rm A_0}$ is not zero in the Zn-free sample with $x=0.115$, implying that the incommensurate magnetic peaks observed from the elastic neutron scattering measurements in the Zn-free sample with $x\sim0.115$~\cite{katano,suzuki,kimura2} do not necessarily correspond to the magnetic order formed throughout the sample. 
These results support our speculation that the dynamical spin correlation or the dynamical stripe correlations tend to be pinned and stabilized around Zn. 

Next, we discuss the peculiar dependence of the magnetic transition on the Zn concentration, as shown in Fig. \ref{fig:a0}. 
In the Zn-free or slightly Zn-substituted samples, the magnetic transition is rather broad, though $T_{\rm N}^{\rm onset}$ is comparatively high. 
With increasing $y$, on the contrary, $T_{\rm N}^{\rm onset}$ decreases as also shown in Fig. \ref{fig:tn}(b) and the transition tends to be sharp. 
This means that the fluctuation of the magnetic transition is large in the Zn-free or slightly Zn-substituted samples. 
In the moderately Zn-substituted samples, on the other hand, the fluctuation of the magnetic transition is comparatively small. 
It appears that Zn operates to pin the dynamical spin correlation or the dynamical stripe correlations so as to suppress the fluctuation of the magnetic transition, as suggested from our transport measurements.~\cite{koike3,koike4,adachi} 
The decrease of $T_{\rm N}^{\rm onset}$ with increasing $y$ may be attributed to weakening of the spin correlation due to the spin-dilution effect of the nonmagnetic Zn.

As for the increase in $T_{\rm N}^{\rm mid}$ with increasing $y$ for $y<0.01$, a qualitatively similar behavior has been observed in the three-dimensional (3D) antiferromagnetically ordered region ($x<0.02$) of La$_{2-x}$Sr$_x$Cu$_{1-y}$Zn$_y$O$_4$ by H\"{u}cker {\it et al}.~\cite{hucker}
They have insisted that the localization of holes around the substituted Zn restores the AF correlation between Cu spins, resulting in the increase in $T_N$ with increasing $y$. 
This scenario may be substantially suitable but is a little different from the case of $x\sim$ 0.115. 
Considering that the dynamical stripe correlations of spins and holes exist at $x\sim0.115$, it is likely that the dynamical stripe correlations are pinned and stabilized by Zn, leading to both the localization of holes in the one-dimensional charge domain and the increase in $T_{\rm N}^{\rm mid}$. 

Finally, we argue the Cu-spin state in the heavily Zn-substituted samples with $y>0.02$. 
The $V_{\rm A_2}$ decreases gradually with increasing $y$ for $y>0.02$ and the muon-spin precession is no longer observed at $y>0.03$ in each $x$, meaning that the magnetic order is destabilized with increasing Zn concentration for $y>0.02$. 
At $y=0.10$, almost no fast depolarization of muon spins is observed, indicating that even a spin-glass state is not realized but the spin correlation becomes weak through the introduction of a large amount of Zn due to the spin-dilution effect of the nonmagnetic Zn.~\cite{hucker} 
Furthermore, considering that the 3D AF order survives through the introduction of Zn up to $y\sim 0.4$ in the Zn-substituted La$_2$Cu$_{1-y}$Zn$_y$O$_4$,~\cite{vajk} the destruction of the magnetic order through only the $5-7$ \% substitution of Zn for $x\sim0.115$ may be attributed to the mobile holes released from the possible stripe order.

In addition, it is worthwhile noting that, at $y\ge0.07$ in $x=0.13$, $V_{\rm A_0}$ is 100 \% but superconductivity does not reappear. 
In the Zn-free sample with $x=0.13$, on the other hand, $V_{\rm A_0}$ is 100 \% and superconductivity takes place. 
Simply thinking, the 7 \% Zn substitution causes scattering of holes so frequently that holes tend to be localized, resulting in the destruction of the superconductivity as in the case of conventional superconductors. 
From the resistivity measurements, however, the value of the resistivity in the normal state at 50 K for the 7 \% Zn-substituted sample with $x=0.13$ is only about six times as large as that for the Zn-free sample with $x=0.13$ and as small as that for the Zn-free sample with $x=0.06$ where superconductivity appears at low temperatures.~\cite{yairi}
Therefore, this may suggest that the dynamical spin correlation or dynamical stripe correlations are essential to the appearance of the high-$T_{\rm c}$ superconductivity. 

\section{Summary}
We have investigated effects of the nonmagnetic impurity Zn on the Cu-spin dynamics and superconductivity from the ZF-$\mu$SR and magnetic-susceptibility measurements in La$_{2-x}$Sr$_x$Cu$_{1-y}$Zn$_y$O$_4$ with $x=0.10$, 0.115 and 0.13 changing $y$ finely up to 0.10. 
The $\mu$SR measurements have revealed that, in each $x$, a magnetic order of Cu spins is observed at $y\sim$ 0.0075, while it disappears and Cu spins turn into a fast fluctuating state for $y>$ 0.03. 
It has been found that all the $\mu$SR spectra can be represented by the three-component function with different frequencies of the Cu-spin fluctuations. 
From the magnetic-susceptibility measurements, on the other hand, it has been found that the volume fraction of the superconducting state rapidly decreases through the slight doping of Zn and that its $y$ dependence corresponds to the $y$ dependence of the volume fraction of the fast fluctuating region of Cu spins estimated from the $\mu$SR results. 
From these results, it has been concluded that the non-superconducting region induced around Zn corresponds to the region where the Cu-spin fluctuations slow down. 
In other words, it has been concluded that both the slowly fluctuating region of Cu spins and the magnetically ordered region compete with the superconductivity. 
Our results have been interpreted in terms of the development of the "Swiss cheese" model. 
That is, it is considered that Zn pins the dynamical spin correlation or dynamical stripe correlations and hence destroys the superconductivity around itself. 

\section*{Acknowledgments}
We would like to thank Prof. Y.J. Uemura and Dr. H. Kimura for fruitful discussions. 
This work was supported by a Grant-in-Aid for Scientific Research from the Ministry of Education, Science, Sports, Culture and Technology, Japan, and also by CREST of Japan Science and Technology Corporation.


\begin{references}
\bibitem{tn}
In the ideal case, $A_0=1/3$ in a magnetically ordered state. In some case, however, $A_0$ becomes smaller than 1/3 even in the ordered state as seen in Fig. \ref{fig:a0}. Therefore, we have decided that $A_0$ becomes $A_{0_{{\rm min}}}$, defined as the mean value among $A_0$ where the magnetic order is formed at 2 K, in the magnetically ordered state for each $x$ and defined $T_{\rm N}^{\rm mid}$ as the midpoint temperature in the change of $A_0$.
\bibitem{a1}
In this calculation, we have assumed that all the Cu spins in the $A_1$ region contribute to the formation of an incoherent magnetic order. 
In the case that all the Cu spins in the $A_1$ region fluctuate slowly, the value of $V_{\rm A_1}$ slightly decreases as mentioned in Sec. \ref{discussion}. 
However, the use of Eq. (\ref{eq3}) for the estimation of $V_{\rm A_1}$ is considered to be reasonable, because almost all Cu spins are nearly or completely magnetically-ordered at 2 K below $T_{\rm N}^{\rm mid}$, except for $y=0.10$ in $x=0.10$, $y\ge0.07$ in $x=0.115$ and $y=0$ and $y\ge0.07$ in $x=0.13$.
\bibitem{3d}
In these pictures, the volume fractions are transformed into the two-dimansional region in the CuO$_2$ plane. It is noted that, even if the volume fractions are transformed into the three-dimensional region in a sample, the discussion in the text is not changed qualitatively.
\bibitem{xiab}
Assuming that all the Cu spins in the $A_1$ regions fluctuate slowly, $\xi_{\rm ab}$ is estimated to be $\sim30$ ${\rm \AA}$.
\end{references}
\end{document}